# *I campanili e i fusi orari*

Una storia di sincronizzazione

## Costantino Sigismondi *

**Sunto:** *Commentiamo il cambio nella misura del tempo durante gli ultimi mesi sui nostri campanili con l'eclissi dell'orologio meccanico, di cui l'Italia è la culla, e anche di quello al quarzo da parte degli orologi radiocomandati.*
*Un breve itinerario ci porta dall'Astrario di Jacopo Dondi dell'Orologio a Padova, all'orologio della Chiesa di Nostra Signora del Suffragio a Torino del Beato Francesco Faà di Bruno, al campanile di S. Antonio a Lanciano dell'Ingegner Antonio Cibotti, acceso da Paolo VI il 4 ottobre 1973, fino agli ultimi aggiornamenti in tema di radiosincronizzazione, che uniformano i rintocchi di campane al fuso orario dell'Europa Centrale entro un secondo di accuratezza.*

**Parole Chiave**: Campanili, orologi, sincronizzazione.

**Abstract:** *The Astrarium made by Jacopo Dondi dell'Orologio in Padua is one of the first public horologia in the World, their public utility is also shown through the history of the tower bell of Our Lady of Suffragio in Turin, made by the mathematician Francesco Faà di Bruno. The tower bell of St. Anthony in Lanciano made by Antonio Cibotti was inaugurated by the Pope Paul VI in 1973 through a radio impulse, and now many tower bells are radio synchronized to the Central European Time within a single second of accuracy.*

⁎ ITIS G. Ferraris e ICRANet, Roma; Observatório Nacional Rio de Janeiro; sigismondi@icra.it..

## 1. Orologi come pendoli semplici

Già nel primo numero di questa rivista ho avuto occasione di scrivere su temi di misurazione del tempo: Il primo solstizio d'inverno alla meridiana di Santa Maria degli Angeli in Roma: misure dell'obliquità dell'eclittica nel 1701. Le meridiane sfruttando il moto diurno del Sole attraverso ombre o immagini stenopeiche consentivano di conoscere con precisione l'istante del mezzogiorno vero che veniva diffuso tramite un segnale visivo o una cannonata a tutta la cittadinanza.

Questo segnale serviva essenzialmente per sincronizzare tutti gli orologi privati, che essendo basati sul movimento di componenti meccaniche, erano e sono soggetti al fenomeno della dilatazione termica e conseguente ritardo o anticipo rispetto al tempo astronomico di riferimento.

In altre parole possiamo immaginare tutti gli orologi meccanici come dei pendoli semplici: quando la temperatura si raffredda la lunghezza del pendolo si contrae e il periodo di oscillazione diminuisce, facendo sì che il secondo scandito dall'orologio sia più breve e l'orologio corra in anticipo, mentre quando la temperatura si riscalda accade il contrario.

Questo fenomeno si riscontra anche negli orologi digitali, al quarzo, che hanno al loro interno una frequenza di oscillazione di riferimento, che è sensibile alla variazione termica. Durante la campagna osservativa di una eclissi di sole, quella anulare del 22 settembre 2006 in Guyana Francese, ho potuto verificare che con il calore equatoriale a cui il mio orologio era stato sottoposto, il ritmo del secondo scandito dall'orologio lo aveva portato in ritardo rispetto agli orologi radiocontrollati del centro spaziale di Kourou.

## 2. L'Astrario di Padova: un'icona tolemaica sempreverde

A Padova nella piazza dell'orologio, a due passi dal Duomo e dal Battistero, c'è l'astrario di Jacopo Dondi dell'Orologio. Una la-pide funeraria di Jacopo sta nel Battistero affrescato da Giusto de Menabuoi, rammenta «Sappi, carissimo lettore, che è mia invenzione quell'orologio che segna il tempo ed il vario fluire delle ore dall'alto della torre qui discosta», all'ingresso del battistero del Duomo di Padova.

Questo astrario di piazza dei Signori mostra il moto del Sole attraverso i segni zodiacali e quello della Luna, riportando l'ora del giorno ed alcune posizioni notevoli (trigono, quadratura e 120°) sempre del Sole con i segni zodiacali, costruito da Jacopo Dondi nel 1344, primo orologio pubblico a segnare, oltre a data ed ora, la posizione del Sole nello Zodiaco e le fasi della luna, venne posto in origine sulla torre della Reggia Carrarese prospiciente Piazza Duomo e andò distrutto nel 1390. L'odierno in Piazza dei Signori è una copia perfetta, eseguita cent'anni più tardi.

Una diversa versione di questo strumento è da attribuire a Giovannni de Dondi, figlio di Jacopo e professore allo Studio Patavino, che ha pubblicato anche un *Tractatus Astrarii*, che ne ha consentito varie repliche fino ai tempi moderni: fu lui ad inventare la *machina admirabilis:* lo stupefacente astrario composto di 300 pezzi.

La sincronizzazione di questo orologio avveniva probabilmente in modo periodico: il mezzogiorno locale osservato mediante meridiane (non se ne conoscono di così antiche, ma non se ne può escludere l'esistenza nel XIV secolo) serviva ad aggiustare l'orologio quando andasse fuori tempo.

La sensibilità di 5 minuti, fornita dall'orologio a cubetti rotanti agli angoli del quadrato che circoscrive l'orologio limita di fatto a 5 minuti la precisione dell'orologio astrario di piazza dei Signori. In termini di mezzogiorno locale, attorno al solstizio d'inverno per esempio, cinque minuti di errore si possono accumulare in 11/12 giorni consecutivi, quando il giorno solare vero vale 24 ore e 25 secondi e sposta in avanti l'istante del mezzogiorno vero di 25 secondi al giorno, rispetto al giorno solare medio, concetto di 5 secoli più giovane nato tra gli studi di Simon Newcombe allo US Naval Observatory. Al tempo di Jacopo Dondi il mezzogiorno locale faceva fede, e il fatto che si spostava di 25 secondi in avanti attorno al solstizio d'Inverno era noto già da Tolomeo e spiegato con complicati calcoli di meccanica circolare. Comunque il problema di avere strumenti meccanici più accurati di 25 secondi al giorno non esisteva nel XIV secolo: si cominciò a ad avere solo nel seicento.

E allora le correzioni col mezzogiorno locale dovevano essere fatte con la tabella dei mezzogiorni calcolata da Tolomeo, in cui, ad esempio, i 25 secondi di ritardo dopo il solstizio d'Inverno erano inclusi.

Dobbiamo aspettare il 1884 con la convenzione di Washington sui fusi orari per assistere alla definitiva scomparsa del mezzogiorno locale dall'uso quotidiano, che fino ad allora aveva dominato.

### 3. Il Campanile di Nostra Signora del Suffragio a Torino

Iniziato nel 1866 e progettato dal matematico e beato Francesco Faà di Bruno, il campanile di Nostra Signora del Suffragio/Santa Zita di Torino è la terza guglia della città.

L'altezza significativa dell'edificio è dovuta ad un motivo curioso e prettamente sociale. Il Faà di Bruno voleva evitare che le lavoratrici e i lavoratori della città venissero ingannati sull'orario di lavoro. Calcolò così che un orologio di due metri di diametro collocato sulle quattro facce del campanile a circa 70 metri di altezza, sarebbe stato visibile in gran parte della città e liberamente consultabile da tutti.

Questo fu anche motivo di richieste ed ottenimento di finanziamenti, e ancora oggi questa opera svetta sulla città della Mole Antonelliana e della Sindone.

La particolarità dell'edificio è la sua altezza 83 metri e l'ampiezza della propria base: appena cinque metri, inoltre la struttura sposa tecniche miste di costruzione. La prima parte è a base quadrata ed è realizzata in muratura a mattoni pieni, a metà della struttura è collocata la cella campanaria riportante delle bifore per ciascun lato, realizzata con 32 colonnine di ghisa per favorire il propagarsi del suono, nonché agevolare l'elasticità strutturale e contrastare la resistenza all'aria; la parte superiore è a base ottagonale, riprendendo il prospetto della cupola ed è realizzata con mattoni forati più leggeri. I prospetti sono scanditi da monofore a tutto sesto, due cornici marcapiano e dai quattro quadranti dell'orologio il cui meccanismo è ivi collocato. La guglia è nuovamente realizzata in ghisa ed è sormontata da un angelo dell'Apocalisse

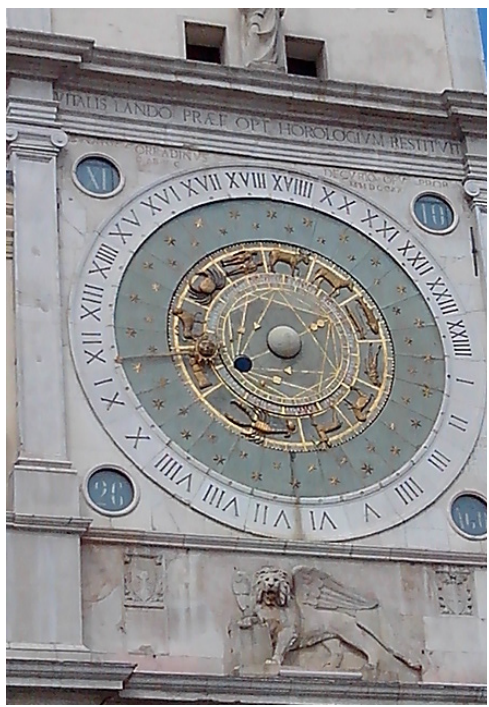

**Fig. 1 - Orologio di Jacopo de Dondi a piazza dei Signori, Padova.**

intento a suonare una tromba.

Ogni campanile aveva questa funzione sociale, anche se oggi è caduta nel dimenticatoio. La stessa scansione del tempo con le campane consente di sapere che ore sono senza neanche guardare l'orologio sulla torre, valendosi della diffrazione dei raggi sonori per la quale giunge alle nostre orecchie il suono della campana anche se la linea di vista è oscurata da ostacoli tra noi e la campana.

Certo è che nella seconda metà dell'ottocento la sincronizzazione dei campanili passa da quella basata sugli orologi solari a quella ibrida in cui il dato dell'orologio solare deve essere integrato da una tabella di longitudine diversa per ogni città, che tiene conto di dove l'orologio solare si trovi collocato nel fuso orario e quale sia la correzione sistematica da apporre al dato astronomico.

### 4. Il Campanile e la radio

Pur essendo oggi il mezzo di sincronizzazione più diffuso, la radio è stata usata per inaugurare un campanile ben prima dell'era degli orologi radiocontrollati. Parliamo del 4 ottobre 1973 quando Paolo VI dal Vaticano accese il campanile di S. Antonio con un segnale radio, come è immortalato nella lapide ai piedi del campanile, che è il più alto di Abruzzo, 72 metri.

Realizzato dall'ingegner Antonio Cibotti, è diventato uno dei simboli di Lanciano, città del Miracolo Eucaristico e del ponte di Diocleziano, che però è visibile da lontano proprio grazie a questo campanile, persino dall'aereo e dalla Majella, montagna situata ad oltre 30 km dalla città e a più di 2000 m di quota.

Chi scrive ha potuto intendere con chiarezza il rintocco delle campane di S. Antonio proprio dalla Majelletta lo scorso 2 Novembre, nei toni più gravi, con il ritardo di quasi 2 minuti dall'istante di diffusione del segnale dal campanile dovuto allo spazio percorso dal segnale.

Questa distanza tanto grande mostra la portata limite teorica delle campane, pensate proprio per la pubblica utilità.

La "Sigismonda" di Cracovia suonava le ore più gravi ed importanti della Polonia, a decine di chilometri di distanza, in una civiltà senza rumori terzi... ero a Niepolomice, nei pressi di Cracovia, quando la Sigismonda suonò nella notte per ricordare i 70 anni dall'inizio della seconda guerra mondiale per la Polonia: ed ho sentito alcuni testimoniare di averla sentita. L'esperienza fatta sulla Majella avendo anche la linea di vista libera mi ha fatto associare quel campanile a quelle campane udite in modo così inatteso in vetta alla Montagna, ricordandomi della "Sigismonda".

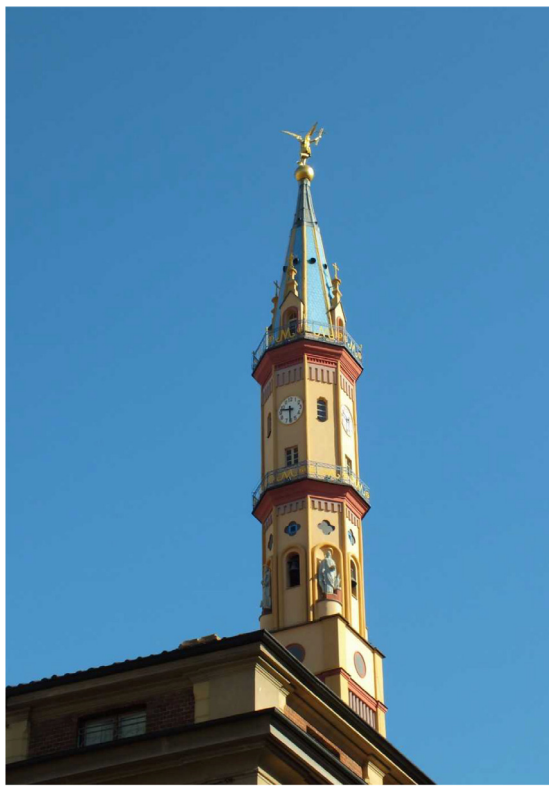

**Fig. 2** - Campanile di N. Signora del Suffragio a Torino.

Da alcuni mesi a questa parte la sincronizzazione del campanile di S. Antonio e quella di altri campanili è passata dalla modalità "a mano" a quella "radiocontrollata".

Questo lo ho potuto notare poiché tengo sempre un orologio sotto controllo per scopo osservazioni astronomiche e dal mese di aprile 2014 ho proprio un modello radiocontrollato della CASIO G-Schock, che si sincronizza ogni notte a mezzanotte.

Per esperienza so che una sincronizzazione al giorno può garantire l'accuratezza entro 0.2 s circa, e so anche che per scopi astrometrici questa precisione non è la massima raccomandabile, tuttavia è sufficiente per capire se gli orologi intorno a me sono sincronizzati nello stesso modo, o possono passare dei mesi prima che qualcuno a mano li risistemi.

Devo dire che alcuni anni fa per me era normale sentire campanili che squillavano in modo casuale entro 4 o 5 minuti attorno all'istante dell'ora intera.

Era una caratteristica di umanità che rendeva piacevole la non sincronizzazione degli orologi... ma oggi questa caratteristica si sta perdendo.

L'arrivo del freddo intenso con impulso di aria polare ha fatto anticipare tutti i campanili della zona dove scrivo, a Roma, di quasi 10 secondi. Può darsi che questa sia l'effetto della meccanica degli orologi che pur radiocontrollati ogni 24 ore abbiano almeno 12 ore di tempo per disallinearsi fino a 10 secondi con circa 1 secondo all'ora di sfasamento termico-meccanico.

Nei giorni di temperatura normale il mezzogiorno veniva scandito con precisione entro il secondo... cosa che non accadeva mai prima d'ora. Sia a Roma che a Lanciano.

Ci sono nuovi standards di puntualità che nemmeno la meccanica può più limitare... l'orologio radiocontrollato è arrivato fino sui campanili delle chiese, nei computer e negli smartphones... prima ero il solo a sapere di questi limiti meccanici, ora il radiocontrollo cancella questo aspetto manuale della realtà condannandoci ad una puntualità atemporale.

Anche le campanelle delle scuole stanno raggiungendo questo standard, sincronizzazione perfetta...ormai solo la relatività ci può liberare dal concetto di tempo che scorre uguale dappertutto, è vero che questo avviene quando le velocità si avvicinano alla velocità della luce, ma la simultaneità vera, di cui questi orologi radiocomandati sembrano essere servi, nella realtà fisica non esiste!

## Referenze

- [http://it.wikipedia.org/wiki/Chiesa_di_Nostra_Signora_del_Suffragio_e_Santa_Zita](http://it.wikipedia.org/wiki/Chiesa_di_Nostra_Signora_del_Suffragio_e_Santa_Zita)
- [http://it.wikipedia.org/wiki/Jacopo_Dondi_dell%27Orologio](http://it.wikipedia.org/wiki/Jacopo_Dondi_dell%27Orologio)
- [http://ricerca.gelocal.it/mattinopadova/archivio/mattinodipadova/2006/01/29/MC6PO_MC601.html](http://ricerca.gelocal.it/mattinopadova/archivio/mattinodipadova/2006/01/29/MC6PO_MC601.html)
- Pier Luigi Bassignana, Faà di Bruno, Torino, Edizioni del Capricorno, 2008.

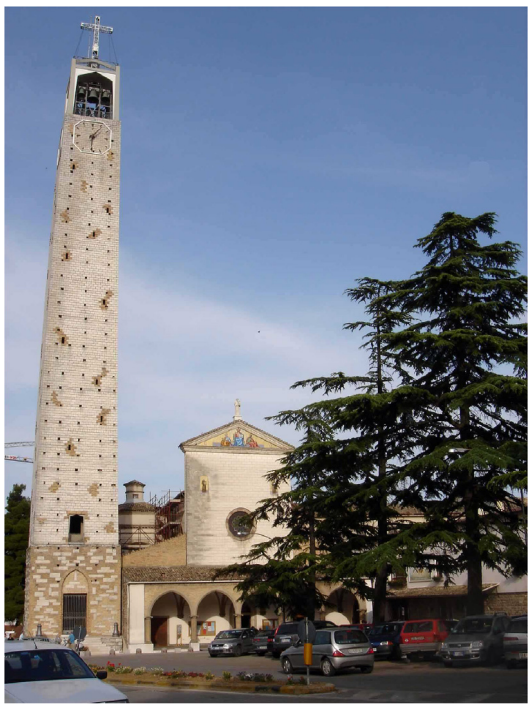

**Fig. 3 -** Campanile e Chiesa di S. Antonio a Lanciano.